\documentclass[aps,prl,amsmath,amssymb,twocolumn,a4paper,superscriptaddress]{revtex4-1}
\usepackage[flushleft]{threeparttable}
\usepackage[T1]{fontenc}
\usepackage[utf8]{inputenc}
\usepackage{graphicx,dcolumn,booktabs,multirow,bigdelim,tabularx,hyperref,multirow}
\usepackage{epstopdf}
\usepackage{xcolor} 
\usepackage{wrapfig} 
\usepackage{sidecap} 
\usepackage{subcaption}

\newcolumntype{Y}{>{\centering\arraybackslash}X}
\begin{document}	

\title{Coulomb explosion of CD$_3$I induced by single photon deep inner-shell ionisation}

\author{M. Wallner}
\affiliation{Department of Physics, University of Gothenburg, 
Origov\"agen 6B, 412 58 Gothenburg, Sweden}

\author{J.H.D. Eland}
\affiliation{Department of Chemistry, Physical and Theoretical Chemistry Laboratory, Oxford University, South Parks Road, Oxford OX1 3QZ, United Kingdom} 
\affiliation{Department of Physics, University of Gothenburg, 
Origov\"agen 6B, 412 58 Gothenburg, Sweden}

\author{R.J. Squibb}
\affiliation{Department of Physics, University of Gothenburg, 
Origov\"agen 6B, 412 58 Gothenburg, Sweden}

\author{J. Andersson}
\affiliation{Department of Physics, University of Gothenburg, 
Origov\"agen 6B, 412 58 Gothenburg, Sweden}

\author{A. Hult Roos}
\affiliation{Department of Physics, University of Gothenburg, 
Origov\"agen 6B, 412 58 Gothenburg, Sweden}

\author{R. Singh}
\affiliation{Department of Physics, University of Gothenburg, 
Origov\"agen 6B, 412 58 Gothenburg, Sweden}

\author{O. Talaee}
\affiliation{Department of Physics, University of Gothenburg, 
Origov\"agen 6B, 412 58 Gothenburg, Sweden}
\affiliation{Nano and Molecular Systems Research Unit, University of Oulu, P.O. Box 3000, FI-90014 University of Oulu, Finland}

\author{D. Koulentianos}
\affiliation{Department of Physics, University of Gothenburg, 
Origov\"agen 6B, 412 58 Gothenburg, Sweden} 
\affiliation{Sorbonne Universit\'e, CNRS, Laboratoire de Chimie Physique-Mati\`{e}re et Rayonnement, F-75005 Paris Cedex 05, France}

\author{M.N. Piancastelli}
\affiliation{Sorbonne Universit\'e, CNRS, Laboratoire de Chimie Physique-Mati\`{e}re et Rayonnement, F-75005 Paris Cedex 05, France}
\affiliation{Department of Physics and Astronomy, Uppsala University, Box 516, SE-751 20 Uppsala, Sweden} 

\author{M. Simon}
\affiliation{Sorbonne Universit\'e, CNRS, Laboratoire de Chimie Physique-Mati\`{e}re et Rayonnement, F-75005 Paris Cedex 05, France}

\author{R. Feifel}
\email{raimund.feifel@physics.gu.se}
\affiliation{Department of Physics, University of Gothenburg, 
Origov\"agen 6B, 412 58 Gothenburg, Sweden} 

\date[]{}

\begin{abstract}
L-shell ionisation and subsequent Coulomb explosion of fully deuterated methyl iodide, CD$_3$I, irradiated with hard x-rays has been examined by a time-of-flight multi-ion coincidence technique.  The core vacancies relax efficiently by Auger cascades, leading to charge states up to 16+. The dynamics of the Coulomb explosion process are investigated by calculating the ions' flight times numerically based on a geometric model of the experimental apparatus, for comparison with the experimental data. A parametric model of the explosion, previously introduced for multi-photon induced Coulomb explosion, is applied in numerical simulations, giving good agreement with the experimental results for medium charge states. Deviations for higher charges suggest the need to include nuclear motion in a putatively more complete model. Detection efficiency corrections from the simulations are used to determine the true distributions of molecular charge state produced by initial L1, L2 and L3 ionisation.
\end{abstract}

\maketitle	


Molecules exposed to sufficiently energetic photons may be totally destroyed in a process where all bonds are broken and all or most of the atoms become positively charged and repel each other by the Coulomb force. This rapid repulsion and subsequent fragmentation of a molecule was termed Coulomb explosion in 1966 by T.A. Carlson and R.M. White \cite{carlsonwhite}. Since then, Coulomb explosions have been studied in many different ways, for instance, by photoion-photoion coincidence spectroscopy in the vacuum ultraviolet (VUV) and soft X-ray region \cite{eland93,multiplychargedICN,molphotodis}, by coincidence imaging \cite{CEI}, by time-resolved pump-probe techniques involving ultraviolet (UV) \cite{uvprobe}, and by X-ray free electron laser (XFEL) ionisation utilising few-photon absorption processes \citep{femtosecondrespons,ultrafastcoulomb,chargeandnuclear,femtosecondcharge,ultrafastdynamics,ultrafastcharge}. The dynamics of the charge rearrangement essential for the Coulomb explosion have previously been studied using time-resolved pump-probe techniques on molecular iodine \cite{electronrearrangement,iodinedynamics}  and in methyl iodide \cite{chargetransfer,imagingcharge} where the results match well with a classical over-the-barrier model.

In this work we have used a multi-ion coincidence technique in combination with hard X-rays to investigate single photon-induced Coulomb explosion of fully deuterated methyl iodide, CD$_3$I, upon creation of a core vacancy in the $n=2$ shell of iodine, building on the original work of Carlson and White \cite{carlsonwhite}, where Coulomb explosion was first explored experimentally by a coincidence method. For the data interpretation, we explore a two-parameter model recently introduced by Motomura et al. \cite{chargeandnuclear} , who studied Coulomb explosion of CH$_3$I from ionisation induced by the absorption of several X-ray photons within a pulse duration of $\sim$ 10 fs.

\begin{SCfigure*}
\includegraphics[width=0.7\textwidth]{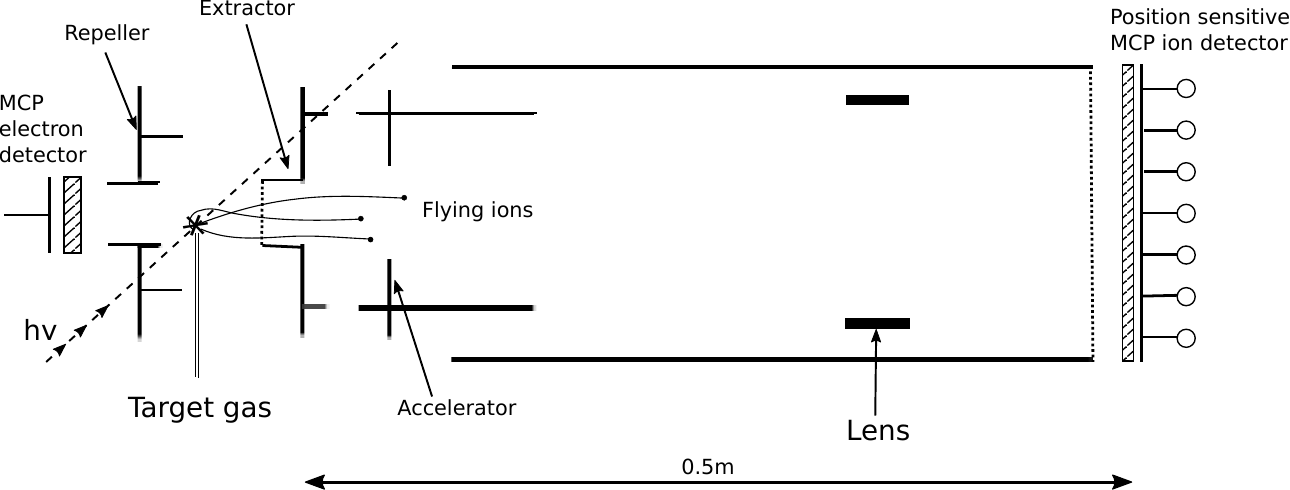}
\caption{Schematic diagram of the multi-ion coincidence spectrometer used. The electrodes around the source region are generating an electric field that accelerates the ions through the aperture of the extractor plate. The ions travel through the drift tube and hit a MCP detector where the time-of-flight is registered.}
\label{fig:expsetup}
\end{SCfigure*}

In the present experiment, photoionisation occurs primarily in selected L-shells of iodine induced by a single photon followed by subsequent Auger cascades, leading to high charge states of CD$_3$I. We measure the time of flight of the ions created in a multi-ion coincidence apparatus, shown schematically in figure \ref{fig:expsetup}, which is described in more detail in the section on experimental methods. After leaving the source volume, a sideways-flying ion of sufficiently high energy will hit the extractor plate rather than pass through its aperture, so the number of ions being detected decreases with increasing kinetic energy. This means that for higher molecular charge states, whose fragment ions gain more kinetic energy, a larger proportion of ions are lost. To quantify the results from our spectrometer in the presence of these effects a numerical investigation was performed. In particular, numerical calculations are used to determine the collection efficiency for each fragmentation channel D$^+$ + C$^{n+}$ + I$^{m+}$. The collection efficiencies are subsequently used to correct the coincidence intensities so that the total number of created events can be determined. Low-order (two-fold and threefold) coincidences are used in most of the analysis to take advantage of their relatively favourable statistics. More details on the implementation of our numerical investigation can be found in the section on numerical methods.

To simulate our experimental data numerically, molecules are  placed within the source volume with a random orientation and are then dissociated and ionised. To determine the mutually dependent angles and kinetic energies with which the ions leave the source, the equation of motions are integrated under mutual Coulomb repulsion. Initially, the atoms start in the normal equilibrium configuration of the molecule, with exact C$_{3v}$ symmetry, making all deuterium atoms equivalent. Time dependence of the ionic charges can, in principle, be calculated by solving a large set of rate equations involving all Auger transition probabilities \cite{rateeq} where a Monte Carlo type approach is imperative \cite{montecarlo}. However, a comparatively simple model introduced recently by Motomura et al. \cite{chargeandnuclear} is computationally more attractive, because it condenses the large set of parameters to two generalised parameters that describe the charge build up and the charge reconfiguration processes simultaneously. The charge is assumed to be sequentially built up, here predominantly by Auger cascade, at the site of the iodine atom, because of its dominant cross section for photon absorption. In this model, the  total charge $Q_{\mathrm{tot}}$ is supposed  to build up according to
\begin{equation}
\label{eq:qtot}
Q_{\mathrm{tot}}(t) = (m + n + 3)\left( 1 - e^{ -t/ \tau } \right) \; ,
\end{equation} 
where $\tau$ is a parameter for the charge build up time, and $m$ and $n$ are the final charges for iodine and carbon, respectively. When a high charge is created on the iodine atom the molecule becomes unstable due to the charge imbalance and electrons from the deuterium atoms and the carbon atom are transferred to the iodine. The rate of transfer is described by
\begin{equation}
\frac{d}{dt} Q_{\mathrm{CD}_3}(t) = R \cdot Q_\mathrm{I}(t) \; ,
\end{equation}
where $R$ is a rate constant for the charge transfer. $Q_I$ and $Q_{CD_3}$ are the charges at the iodine site and the methyl group, respectively, and they obey
\begin{equation}
Q_{\mathrm{tot}}(t) = Q_\mathrm{I}(t) + Q_{\mathrm{CD}_3}(t) \; .
\end{equation}
The ions are allowed to have fractional charges during the charge build up, which is assumed by considering effects such as delocalisation and screening, but are all required to have integer final charges. Charges of at least 4+ on the methyl group are apportioned as three units to the three equivalent  deuterium atoms and the residue to the carbon atom. For comparison with the charge build up model, we also calculate the result of instantaneous creation of the final charges on the atoms in their original positions.

\begin{figure*}
\includegraphics[width=\textwidth]{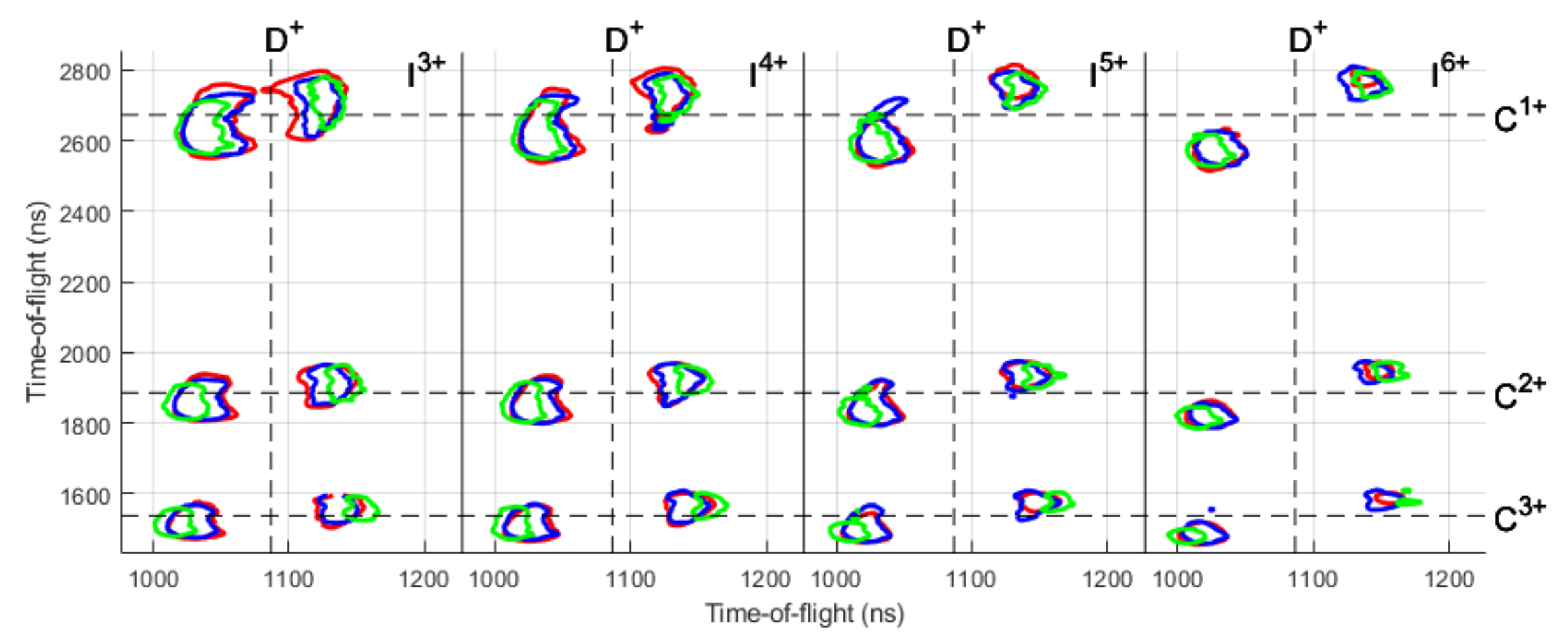}
\caption{Contours of deuterium-carbon ion pairs of charges 1+ and n+, respectively, correlated with different charges m+ of iodine. The vertical axis shows the carbon separation and the horizontal axis shows the deuterium separation. Dashed lines represent the flight times for zero kinetic energy ions. The red contours are the experimental data, the blue are the numerical data based on the charge build up model with charge transfer and the green are the numerical data for an instantaneous model. The contour lines are for 6\% of maximum intensity.}
\label{fig:coincomp}
\end{figure*}

\begin{figure}
\includegraphics[width=0.5\textwidth]{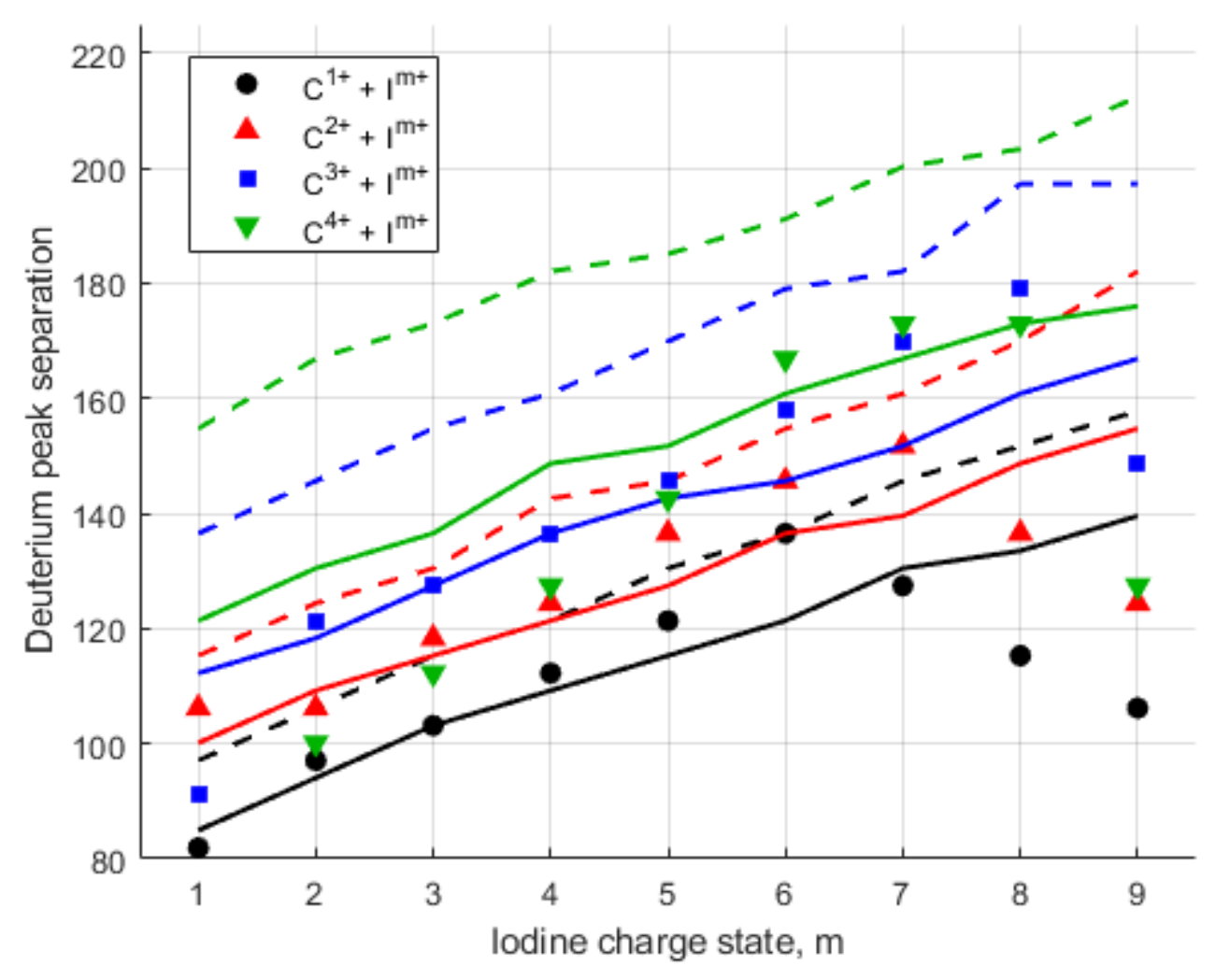}
\caption{Comparison of the islet peak separation of the deuterium ion involved in triple coincidence events with C$^{n+}+$I$^{m+}$. Markers: experimental data; dashed lines: instantaneous model; solid lines: charge build-up model.}
\label{fig:peaksep}
\end{figure}


Using the model described above in comparison with our experimental results, starting from the parameters in \cite{chargeandnuclear}, we find the best overall agreement with $\tau = 7$ fs and $R = 0.37$ fs$^{-1}$. However, varying $R$ has only to a minor effect on the kinetic energy release and therefore the simulated flight times. The parameters are determined by comparing the experimental and numerical time-of-flight distributions from triple coincidence detections from different fragmentation channels D$^+ +$C$^{n+}+$I$^{m+}$. The lighter ionic species are the most sensitive to change in the model parameters; in particular the deuterium ion is the most suitable for comparison between the model's predictions and the experimental data. To ensure that a specific decay channel is involved in each comparison it is required that iodine and carbon ions are detected in addition to one deuterium ion. Triple coincidences are used as the numbers of fourfold and fivefold coincidences are very low. The simulations suggest that only a very small set of initial molecular orientations lead to trajectories allowing detection of more than one D$^+$ ion. Fig. \ref{fig:coincomp} shows a comparison between the experiment and the simulations in form of triple detections showing the ion pair contours of deuterium-carbon ions of charges 1+ and n+, respectively, together with different iodine species. The figure contains experimental data in red contour lines, blue contours show the numerical data utilising Motomura's model \cite{chargeandnuclear} with the previously mentioned parameters, and the green contour lines show numerical data using the instantaneous model. The ion pairs form two islets, which is a consequence of the high momenta gained which result in sideways flying ions missing the detector. As evidenced by the figure, the charge build up model for these intermediate decay channels, n $\in[1,3]$ and m $\in[3,6]$, matches well with the experiment whereas the instantaneous model overestimates the kinetic release energies, especially for the deuterium fragment. For a comprehensive reflection of all the resolvable decay channels, an islet separation plot is shown in Fig. \ref{fig:peaksep}, showing the separation of the deuterium 1+ ion correlated with carbon and iodine ions with positive charges of 1-4 and 1-9, respectively. The islet separation is selected as the time at half the full peak height on the outside of each peak (forward and backward), and the figure shows the experimental data in symbols, the data based on the numerical charge build up model in solid lines and the data based on the instantaneous model in dashed lines. The figure shows that the model recreates Coulomb explosion for the intermediate decay channels in good agreement with the experiment, whereas decay channels involving a high carbon or iodine charge deviate and the fragments gain less momenta from Coulomb explosion. The deviations in low iodine charges for carbon 4+ may be explained by the lack of charge imbalance in favour of the iodine ion, as assumed by the charge build up model, when in fact the carbon ion has a higher charge during the majority of the charge build up.

In order to determine the true charge distributions produced by each specific initial hole creation, we use the relative intensities of threefold coincidences for each channel. Although only one D$^+$ ion is usually detected all three are assumed to become charged in the explosions and this assumption underlies all the derived distributions. Several corrections must be applied to the raw intensities to account for differences in ion and electron detection probabilities.

\begin{figure}[b]
\includegraphics[width = 0.5\textwidth]{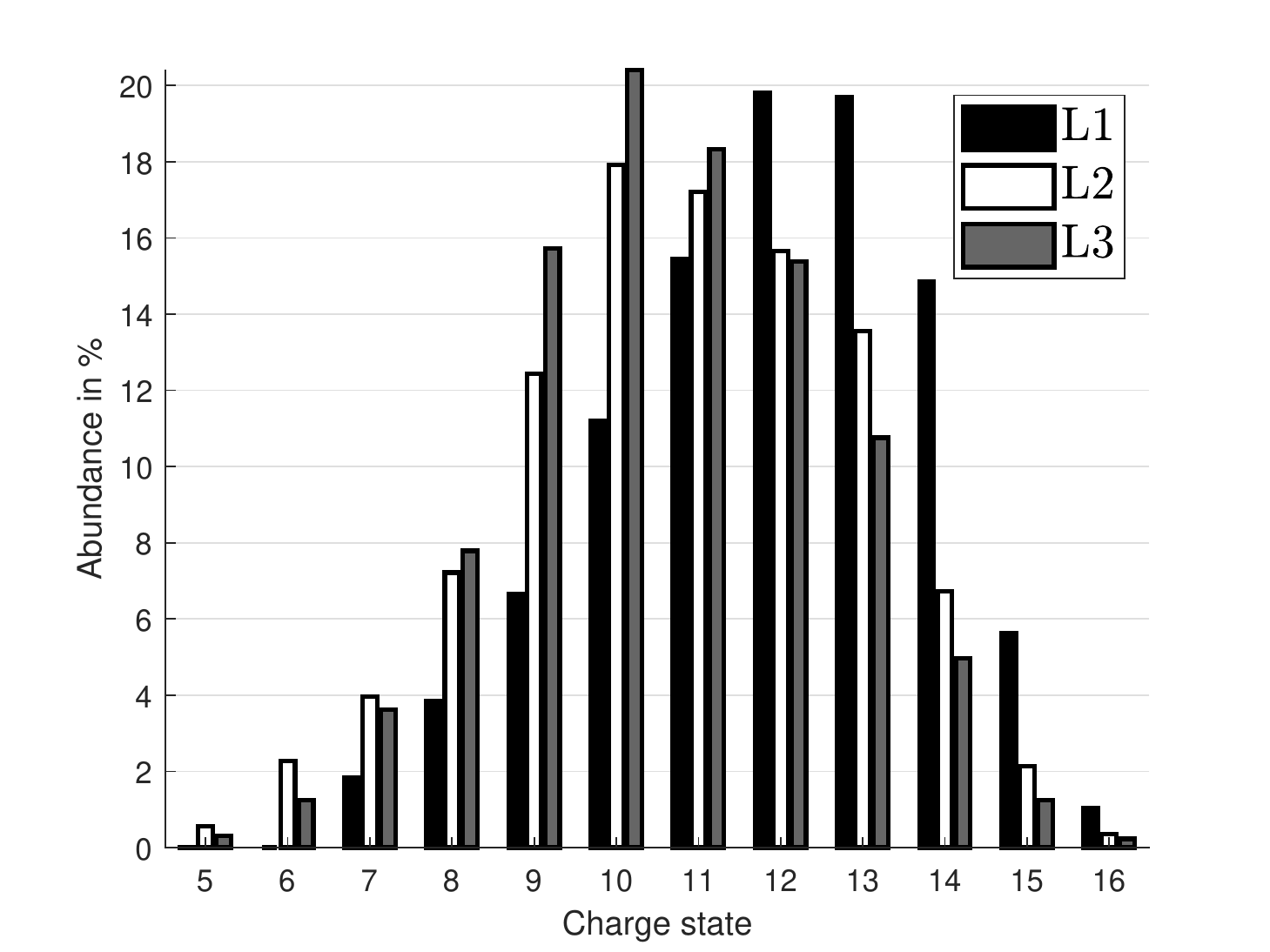}
\caption{Relative charge state abundances estimated from the experimental data, and corrected for the simulated collection efficiency. The bars reflect the abundance of the charge states present in initial ionisation from the iodine 2s, 2p$_{1/2}$ and 2p$_{3/2}$ shells, represented in black, white and grey, respectively.}
\label{fig:abundance}
\end{figure}

Because of the geometry of the apparatus shown in Fig. \ref{fig:expsetup}, fragmentation releasing higher initial kinetic energy implies a greater loss of sideways flying ions. The collection efficiency is non-linear as a function of kinetic energy and is different for each ionic species and consequently the fragmentation channels are affected differently. In order to determine the true distribution of molecular charge states this non-linear dependence in the apparatus is determined numerically through simulation. Another factor is the risk of not detecting any electrons to initialise a flight time measurement. The probability of detecting a single electron is $f_i \approx  50\%$ and thus the probability of not detecting any electrons for $n$ ejected electrons is  $(1-f_i)^n$. In the present case, the lowest resolvable molecular charge state involves 5 emitted electrons, yielding a  probability of not detecting any start electron of $\sim 3\%$.

Experimental charge distributions are extracted at four different photon energies near the L-subshell thresholds in iodine such that each set of data  contains a different blend of ionisation events from the different subshells. A subtraction method is implemented to determine the distributions of charge states from the individual L1, L2 and L3 ionisations, on the basis of the theoretical prediction that the total photoelectric cross-section \cite{photoelectriccrosssection} for each ionisation declines as a function of photon energy, $E$,  in line with a fitted polynomial, with $E^{-7/2}$ as the dominating term \cite{berkowitz}. The four data sets are denoted as $D1-D4$, where $D1$ corresponds a to photon energy of 5290 eV which is above L1, L2 and L3 thresholds, $D2$ to 4950 eV which is above the L2 and L3 thresholds, $D3$ to 4660 eV which is only above the L3 threshold, and $D4$ at 4300 eV which is below all the L threshold and shows ionisation only from shells at lower binding energies.

Using the theoretical relative partial cross-sections derived as above, we obtain a set of equations allowing extraction of the charge distributions produced by each pure subshell ionisation: \begin{align}
    \text{L3} &= D3 - 0.3251\cdot D4 \\
    \text{L2} &= D2 - 0.7518\cdot D3 + 0.0125\cdot D4 \\
    \text{L1} &= D1 - 0.8076\cdot D2 - 0.0704\cdot D3 + 0.0121\cdot D4 \;.
\end{align}

where the data set $D4$ has considerably poorer statistics than the other three. However, the uncertainty of $D4$ is a major factor only in the  extraction of the pure L3 distribution. The distribution for pure L1 has a relatively large uncertainty as the subtracted terms constitute a large fraction of the D1 data set. The distributions of total molecular charge produced by ionisation from the individual shells are shown in Fig. \ref{fig:abundance}, which shows that  the most probable charge number for L1 ionisation is roughly two units higher than those from  L2 and L3 ionisation. This can be related to a fast Coster-Kronig transition where the L1 hole is filled by an L3 electron, providing enough energy to eject an electron from the M shell after which the relaxation proceeds similarly to the case where the initial hole is created in L3, with an additional M hole. A similar transition for an initial L2 hole is expected to be much less probable as the energy available from an L3 to L2 transition is only sufficient to eject an N electron and the increased radial distance limits the transition rate.

The present study shows that the charge creation model introduced by Motomura et al. \cite{chargeandnuclear} using only a few empirical parameters gives a satisfactory description of single photon induced Coulomb explosion for the decay channels involving low and intermediate charges. But the discrepancy for decay channels involving higher charge states of carbon and iodine indicates that a better model may need to incorporate  specific effects of nuclear motion on charge development and/or charge transfer when ionisation is by deep hole creation followed by Auger cascade. A fuller analysis may also need to allow for the possible involvement of neutral fragments in the Coulomb explosion. 

A related effect ignored in the present model is the possible role played by molecular vibrational modes. The energetic contribution of vibrational modes is most likely negligible, but some degenerate modes deform the molecule in such a way that the most probable structure at any single instant is different from the time-averaged (C$_{3v}$) structure. Such a deformation may affect the Coulomb explosion of CD$_3$I by making the deuterium atoms inequivalent  at the instant of ionisation. The normal mode expected to be dominant in the zero-point motion of such a study is the $\nu_6(e)$ mode. The present experiments were unfortunately not sensitive enough to investigate such effects, but they should be detectable  in future experiments with fully operational ion detection  position sensitivity.


In  conclusion, we have investigated single photon induced Coulomb explosion of fully deuterated methyl iodide, CD$_3$I, using X-ray pulses in the 4660-5290 eV photon energy region. Our experimental ion time-of-flight data were compared to the results of a numerical model which took into account the dimensions of our apparatus and the electrical fields applied. The few-parameter  model used here  has previously been shown to work well in a Coulomb explosion study of CH$_3$I induced by multi-photon ionisation involving an XFEL. The present study shows that this model is also applicable to the single photon case for intermediate charge state decay channels, as evidenced by the good agreement between our experimental and numerical results. Decay channels involving high carbon or iodine charge show a systematically lower  kinetic energy release than predicted by the model, which we attribute to a stronger effect of competitive nuclear motion.  If our interpretation is correct, for the light hydrogen isotopologue, CH$_3$I, stronger deviations in kinetic energy release from the model predictions may be expected.  The success of the simple model is perhaps somewhat surprising in view of the very different charge generation pathways in deep hole creation and Auger cascade as opposed to x-ray multi-photon ionisation.  The model may serve as a useful tool for description of Coulomb explosions more generally, and as a starting point for  more sophisticated modeling.  
Experimental abundances together with detection efficiencies derived from numerical modeling of the apparatus were used to determine the true relative abundances of  initial charge states. The charge states from L2 and L3 ionisation have similar distributions, whereas L1 ionisation gives a distribution displaced roughly two units higher in charge. This difference is explained by a rapid Coster-Kronig transition turning an L1 electron hole into an L2 or L3 electron hole with an additional electron hole in a higher shell.

\section{Experimental method}

The experiments were carried out using synchrotron radiation provided by the LUCIA beam line of the storage ring SOLEIL in Paris. The ring was operated in single bunch mode at a frequency of $0.83$ MHz. The synchrotron radiation pulses interacted in a crossed-beam configuration with an effusive jet of the target gas from a hollow needle located in the source region of the apparatus shown in Fig. \ref{fig:expsetup}. The electrodes provide an electric field that accelerates the negative electrons to a nearby detector, serving as a start for the ion coincidence measurements, while the positive ions fly in opposite direction to a more distant, position sensitive microchannel plate detector \cite{mcp}. The position information on the data hits has not been used in the analysis because of a strong lensing effect which essentially guided almost all ions to a small area in the centre. Some angular effects are preserved, but they are too small to be interpreted reliably.

Since the ionic fragments created by the Coulomb explosion may gain a fairly high kinetic energy release, not all of the ions reach the detector. In practice, only about 21\% of the deuterium atoms, 50\% of the carbon atoms and 78\% of the iodine atoms reach the detector. This is due to the geometrical configuration of the apparatus where sideways flying ions of sufficient energy will hit the inside of the extractor plate rather than passing through the aperture.
The loss of sideways flying ions create a hollowness in the overall time-of-flight peak shapes, yielding a forward- and a backward component separated by $\frac{2\sqrt{2m U_0}}{qE}$.

Ions that do reach the detector may still not be registered because of the non-unit efficiency of the detector which is predominantly determined by the open area ratio of the channel plate detector.

\section{Numerical methods}
In the model of the Coulomb explosion process used, the molecule is assumed to start at its nominal equilibrium geometry with exact C$_{3v}$ symmetry, representing all deuterium atoms equivalently. A time dependent charge is given to the atoms as they are allowed to evolve, numerically calculated using ode45 in Matlab, under their mutual Coulomb repulsion until they have moved sufficiently far apart that their energies and relative angles are fixed. At this point the energies and angles are transformed into LAB coordinates with a random orientation and the further motion of the ions is determined by the applied electric fields.

A geometric model of the apparatus was constructed in SIMION \cite{simion}, to mimic the experimental conditions as realistically as possible. The initial positions, kinetic energies and angles of all ions from each explosion determined by the simulation are fed into this model of the apparatus, where they travel towards the detector under the influence of the applied electric fields. This allows the fragments of a molecule to be modelled as coincidence events, incorporating the  loss of sideways-flying ions to allow comparison with the actual experiment. The modelling takes into account the limited efficiency of the detector by random deletion of half the ions that arrive at the detector surface. Once the simulated data resemble the experimental data set sufficiently closely the initial charge state abundances can be extracted.

\section{Author information}

\subsection{Author contribution}
J.H.D.E., M.N.P., M.S. and R.F. devised the research,  R.J.S., O.T., R.S., J.A., A.H.R., D.K., and R.F. participated in the conduction of the experimental research, M.W. constructed the numerical model, M.W. and J.H.D.E. performed the data analysis, M.W., J.H.D.E. and R.F. wrote the paper and all authors discussed the results and commented on the manuscript at several instances.

\subsection{Notes}
The authors declare no competing interests.

\section{Data availability}
The datasets generated during and/or analysed during the current study are available from the corresponding author on reasonable request.

\section{Acknowledgement}

This work has been financially supported by the Swedish Research Council (VR) and the Knut and Alice Wallenberg Foundation, Sweden. We thank SOLEIL for the allocation of synchrotron radiation beam time and we want to warmly acknowledge the staff and colleagues of this facility for their technical assistance and administrative support.


\end{document}